# Spatio-temporal data mining in ecological and veterinary epidemiology

Aristides Moustakas

School of Biological and Chemical Sciences

Queen Mary University of London

Mile End Road, London E1 4NS, UK

Aristides (Aris) Moustakas

Email: arismoustakas@gmail.com

**Abstract**
Understanding the spread of any disease is a highly complex and interdisciplinary exercise as biological, social, geographic, economic, and medical factors may shape the way a disease moves through a population and options for its eventual control or eradication. Disease spread poses a serious threat in animal and plant health and has implications for ecosystem functioning and species extinctions as well as implications in society through food security and potential disease spread in humans. Space-time epidemiology is based on the concept that various characteristics of the pathogenic agents and the environment interact in order to alter the probability of disease occurrence and form temporal or spatial patterns. Epidemiology aims to identify these patterns and factors, to assess the relevant uncertainty sources, and to describe disease in the population. Thus disease spread at the population level differs from the approach traditionally taken by veterinary practitioners that are principally concerned with the health status of the individual. Patterns of disease occurrence provide insights into which factors may be affecting the health of the population, through investigating which individuals are affected, where are these individuals located and when did they become infected. With the rapid development of smart sensors, social networks, as well as digital maps and remotely-sensed imagery spatio-temporal data are more ubiquitous and richer than ever before. The availability of such large datasets (Big data) poses great challenges in data analysis. In addition, increased availability of computing power facilitates the use of computationally-intensive methods for the analysis of such data. Thus new methods as well as case studies are needed to understand veterinary and ecological epidemiology. A special issue aimed to address this topic.

*Introduction*

Understanding the spread of any disease is a highly complex and interdisciplinary exercise as biological, social, geographic, economic, and medical factors may shape the way a disease moves through a population and options for its eventual control or eradication (Aristides Moustakas & Evans, 2016; Oleś, Gudowska-Nowak, & Kleczkowski, 2012). Disease spread poses a serious threat in animal and plant health and has implications for ecosystem functioning and species extinctions (Fisher, et al., 2012) as well as implications in society through food security and potential disease spread in humans (Graham, et al., 2008; Tomley & Shirley, 2009).

Space-time epidemiology (Knox & Bartlett, 1964) is based on the concept that various characteristics of the pathogenic agents and the environment interact in order to alter the probability of disease occurrence and form temporal or spatial patterns (Snow, 1855; Ward & Carpenter, 2000). Epidemiology aims to identify these patterns and factors, to assess the relevant uncertainty sources, and to describe disease in the population. Thus disease spread at the population level differs from the approach traditionally taken by veterinary practitioners that are principally concerned with the health status of the individual (Arah, 2009). Patterns of disease occurrence (Markatou & Ball, 2014) provide insights into which factors may be affecting the health of the population, through investigating which individuals are affected, where are these individuals located and when did they become infected.

*Technological advancements*

With the rapid development of smart sensors (Aanensen, Huntley, Feil, al-Own, & Spratt, 2009), social networks, as well as digital maps and remotely-sensed imagery spatio-temporal data are more ubiquitous and richer than ever before (Gange & Golub, 2016) epidemiology in the big data era needs to integrate novel methods (Mooney, Westreich, & El-Sayed, 2015; Pfeiffer & Stevens, 2015). The availability of such large datasets (big data) poses great challenges in data analysis (Fan,

Han, & Liu, 2014; Najafabadi, et al., 2015). In addition, increased availability of computing power facilitates the use of computationally-intensive methods for the analysis of such data (Aristides Moustakas & Evans, 2015). Data mining - methods combining statistics and computer science - are increasingly employed (Lynch & Moore, 2016) and may provide novel insights into epidemiological problems (McCormick, Ferrell, Karr, & Ryan, 2014; Nelson, et al., 2014)

*Let the data speak?*

Can big data replace theory? It has been suggested that the availability of a large volume of data, data deluge will make the scientific method obsolete (Anderson, 2008); hypothesis-driven, or equation-driven research will become irrelevant and data mining will be used instead (Anderson, 2008). This thesis has generated a large scientific discussion – for some examples across scientific disciplines see (Benson, 2016; Chiolero, 2013; Levallois, Steinmetz, & Wouters, 2013; Toh & Platt, 2013), for online discussions see: https://www.edge.org/discourse/the_end_of_theory.html. Adding up to the discussion it has been suggested that experts will decline in importance in the big data sector (Mayer-Schönberger & Cukier, 2013). There are cases where model-free forecasting (using machine learning methods) outperforms the correct mechanistic model for simulated and experimental data (Perretti, Munch, & Sugihara, 2013). However if one simply relies on data-driven science several components of scientific methods will be made poorer: thought experiments (McAllister, 1996), stochastic reasoning (Christakos, 2010; Pearl, 1987), or theoretically-derived predictions may open a new field and propose as a testable hypothesis (Gorelick, 2011); something feasible in the mathematical universe is something that may happen in the biological/physical universe (regardless upon how likely is that to happen). A classic example derives from Einstein's general relativity theory. The theory was based on the observed difference for Mercury's precession between Newtonian theory and observation i.e. the deviance between observation and a model. The theory at the time that was developed lacked data but it was at later time steps verified by data. A data-driven science is welcome but we cannot afford to lose well established, tested through time scientific methods.

*Are more data always better?*

While the answer may look an obvious yes and that the only challenge is how to handle, visualise, and analyse large datasets, this is not always the case. Big datasets bring a lot of spurious correlations which appear to be simply relationships between things that are just random noise (Silver, 2012). In addition big datasets allow easier 'cherry-peaking', people can choose which fractions of the data to use in order to show something that they already support or simply to produce a novel result, while a larger dataset may have simply falsified the reported result (Silver, 2012), or simply verified something that was already known (Donoho & Jin, 2015), therefore this would not merit a groundbreaking result/publication (Silver, 2012). In addition, factor analysis in time series in econometrics showed that collating several datasets together may generate cross-correlated idiosyncratic errors, or a dominant factor in a smaller dataset may be a dominated factor in a larger dataset (Boivin & Ng, 2006). In such cases smaller datasets have yielded results at least as satisfactory or in fact even better than larger datasets (Boivin & Ng, 2006; Caggiano, Kapetanios, & Labhard, 2011). Methods accounting for the effects of cross correlated errors have been proposed (Blair & Bar-Shalom, 1996). While these examples are mentioned in order to highlight problematic issues related with big data, more often than not certainly more data are desirable than fewer.

*Data availability and model complexity*

A study in climate modelling has shown that as the models are becoming increasingly complex and realistic, they are also becoming less accurate because of cumulative uncertainties (Maslin & Austin, 2012). In the case of climate modelling earlier models did not account for many important factors that are now being included (Maslin & Austin, 2012). The simplicity of the models also prevented the uncertainties associated with these factors from being included in the modelling.

The uncertainty remained hidden. More complex models that include more factors are also associated with higher uncertainties (Maslin & Austin, 2012). There is thus the paradox that as models are becoming more complex and more realistic (matching the real world better) they also become more uncertain. Ecological systems are quite complex with many small tapering effects, large heterogeneity, and interactions that are generally unknown. On an information-theoretic approach, 'information' about the biological system under study exists in the data and the goal is to express this information in a compact way (Evans, et al., 2014; Lonergan, 2014); the more data available the more information exists, i.e. a more complicated statistical model may approximate the data (Burnham & Anderson, 2002) and more complex predictive models (process based models such as individual based models) may be calibrated (Evans & Moustakas, 2016).

*The importance of pubic data.*

While several new technologies providing a large volume of data exist (mentioned earlier in this paper), publicly available data from governmental organizations as well as data sharing among scientists (Michener, 2015) having public data repositories are easier than ever due to large computer storage availabilities as well as fast network connections for downloading them. These public data promote transparency and accountability in the analysis, the potential for data expansion by merging several datasets together, as well as building up the impact of the work (Kenall, Harold, & Foote, 2014; Piwowar & Vision, 2013). In order to predict and mitigate disease spread informed decisions are needed. Often decisions involve conflicts between several stakeholders (Krebs, et al., 1998; A. Moustakas, 2016). These decisions need to be taken based on data analysis and predictive models calibrated with data. Making publicly available data will greatly facilitate their analysis and to informed decisions. For a review of publicly available veterinary epidemiological data with web sources links see (Pfeiffer & Stevens, 2015).

*Spatio-temporal Data mining in Veterinary and Ecological Epidemiology*

There is thus a need for new methods as well as case studies to enhance our understanding in spatio-temporal data mining in veterinary and ecological epidemiology. A special issue in the journal Stochastic Environmental Research and Risk Assessment aimed to address this topic. Potential thematics included: spatiotemporal statistics (Biggeri, Catelan, Conesa, & Vounatsou, 2016; Picado, Guitian, & Pfeiffer, 2007), stochastic analysis (Heesterbeek, 2000; Marx, Mühlbauer, Krebs, & Kuehn, 2015), Bayesian maximum entropy modeling (Biggeri, et al., 2006; Juan, Díaz-Avalos, Mejía-Domínguez, & Mateu, in press), big data analytics (Andreu-Perez, Poon, Merrifield, Wong, & Yang, 2015; Guernier, et al., 2016), GIS and Remote Sensing (Ferrè, et al., 2016; Norman, et al., 2012), Trajectories and GPS tracking (Demšar, et al., 2015; Zhang, et al., 2011), Agent Based Modelling calibrated with data (Dion, VanSchalkwyk, & Lambin, 2011; Aristides Moustakas & Evans, 2015; Smith?, et al., 2016), decision making and risk assessment (Fei, Wu, Liu, Ren, & Lou, 2016; Lowe, Cazelles, Paul, & Rodó, in press), network and connectivity analysis (Nobert, Merrill, Pybus, Bollinger, & Hwang, 2016; Ortiz-Pelaez, Pfeiffer, Soares-Magalhaes, & Guitian, 2006) and co-occurrence and moving objects (Miller, 2012; Webb, 2005). Nine contributions were finally accepted after peer reviewing.

Bayesian analysis of spatial data often uses a conditionally autoregressive prior, expresses spatial dependence commonly present in underlying risks or rates. These conditionally autoregressive priors assume a normal density and uniform local smoothing for underlying risks often violated by heteroscedasticity or spatial outliers encountered in epidemiological data. (Congdon, this volume) proposes a spatial prior representing spatial heteroscedasticity within a model accommodating both spatial and non-spatial variation. The method is applied both in a simulation example based on US states, as well as in a real data application considers Tuberculosis incidence in England (Congdon, this volume). The code used for generating simulations is also provided in R (R Development Core Team, 2016).

An understanding of the factors that affect the spread of endemic bovine tuberculosis is critical for the control of the disease. Analyses of data need to account for spatial heterogeneity, or spatial autocorrelation may inflate the significance of explanatory covariates. (Brunton, Alexander, Wint, Ashton, & Broughan, this volume) used three methods, least-squares linear regression with a spatial autocorrelation term, geographically weighted regression, and boosted regression tree analysis, to identify the factors that influence the spread of endemic bovine tuberculosis at a local level in England and Wales. The methods identified factors related to flooding, disease history and the presence of multiple genotypes of endemic bovine tuberculosis and these factors were consistent across two of the three methods (Brunton, et al., this volume).

Early warning indicators are particularly useful for monitoring and control of any disease. (Malesios, Kostoulas, Dadousis, & Demiris, this volume) provide an early warning method of sheep pox epidemic applied in data from Evros region, Greece. To provide inference on the mechanisms governing the progress of sheep pox epidemic (Malesios, et al., this volume) follow a two-stage procedure. At the first stage, a stochastic regression model is fitted to the complete epidemic data. The second stage uses an analogy of the fitted model with branching processes in order to obtain a system of estimating the probability of the epidemic going extinct at each of several time points during this epidemic. The end result is an evidence-based early warning system that could inform the authorities about the potential spread of the disease, in real-time.

Japanese encephalitis, a vector-borne disease transmitted by mosquitoes and maintained in birds and pigs. To examine the potential epidemiology of the disease in the USA, (Riad, Scoglio, McVey, & Cohnstaedt, this volume) use an individual-level network model that explicitly considers the feral pig population and implicitly considers mosquitoes and birds in specific areas of Florida and Carolina. To model the virus transmission among feral pigs, two network topologies are considered: fully connected and random with a defined probability networks. Patterns of simulated outbreaks support the use of the random network similar to the peak incidence of the closely related West Nile virus, another virus in the Japanese encephalitis group (Riad, et al., this volume). Simulation analysis suggested two important mitigation strategies.

Disease outbreaks are often followed by a large volume of data, usually in the form of movements, locations and tests. These data are a valuable resource in which data analysts and epidemiologists can reconstruct the transmission pathways and parameters and thus devise control strategies. However, the spatiotemporal data gathered can be both vast whilst at the same time incomplete or contain errors. (Enright & O'Hare, this volume) provide a user friendly introduction to the techniques used in dealing with the large datasets that exists in epidemiological and ecological science and the common pitfalls that are to be avoided as well as an introduction to Bayesian inference techniques for estimating parameter values for mathematical models from spatiotemporal datasets. The analysis is showcased with a large dataset from Scotland and the code and data used in this paper are also provided (Enright & O'Hare, this volume).

Mechanistic epidemiological modelling has a role in predicting the spatial and temporal spread of emerging disease outbreaks and purposeful application of control treatment in animal populations. (Lange & Thulke, this volume) address the newly emerging epidemic of African swine fever spreading in Eurasian wild boar using an existing spatio-temporally explicit individual-based model of wild boar. (Lange & Thulke, this volume) propose a mechanistic quantitative procedure to optimise calibration of several uncertain parameters based on the spatio-temporal simulation model output and the spatio-temporal data of infectious disease notifications. The best agreement with the spatio-temporal spreading pattern was achieved by parameterisation that suggests ubiquitous accessibility to carcasses but with marginal chance of being contacted by conspecifics e.g., avoidance behaviour. The parameter estimation procedure is fully general and applicable to problems where spatio-temporal explicit data recording and spatial-explicit dynamic modelling is performed.

In the last two decades, two important avian influenza viruses infecting humans emerged in China, the highly pathogenic avian influenza H5N1 virus, and the low pathogenic avian influenza

H7N9 virus. China is home to the largest population of chickens and ducks, with a significant part of poultry sold through live-poultry markets potentially contributing to the spread of avian influenza viruses. (Artois, et al., this volume) compiled and reprocessed a new set of poultry census data and used these to analyse H5N1 and H7N9 distributions with boosted regression trees models. (Artois, et al., this volume) found a positive and previously unreported association between H5N1 outbreaks and the density of live-poultry markets.

Transmitted infectious diseases, aggregate regional chronic diseases, and seasonal or transitory acute diseases can cause extensive morbidity, mortality and economic burden. Since the space–time distribution of a disease attribute is generally characterized by considerable uncertainty, the attribute distribution can be mathematically represented as a spatiotemporal random field model. (Christakos, Zhang, & He, this volume) present a random field model of disease attribute that transfers the study of the attribute distribution from the original spatiotemporal domain onto a lower-dimensionality travelling domain that moves along the direction of disease velocity. The partial differential equations connecting the disease attribute covariances in the original and the travelling domain are derived with coefficients that are functions of the disease velocity. The theoretical model is illustrated and additional insight is gained by means of a numerical mortality simulation study, which shows that the proposed model is at least as accurate but computationally more efficient than mainstream mapping techniques of higher dimensionality (Christakos, et al., this volume).

(Aristides Moustakas & Evans, this volume) use a very large dataset generated by a calibrated agent based model to perform network analysis, spatial, and temporal analysis of bovine tuberculosis between cattle in farms and badgers. Infected network connectedness was lower in badgers than in cattle. The contribution of an infected individual to the mean distance of disease spread over time was considerably lower for badger than cattle. The majority of badger-induced infections occurred when individual badgers leave their home sett, and this was positively correlated with badger population growth rates. The spatial aggregation pattern of the disease in cattle and badgers is different across scales – in badgers, we find that the disease is found in clusters whereas in cattle the disease is much more random and dispersed. There is little geographical overlap between farms with infected cattle and setts with infected badgers, and cycles of infections between the two species are not synchronised. The findings reflect the movements of the animals – for example, cattle move greater distances within their grounds or they can be sold to farms further afield. Conversely, badgers are social animals that live in groups, and rarely leave their homes, meaning that the presence of TB is more clustered (Christakos, et al., this volume). The research suggests that an efficient way to vaccinate badgers might be to follow the spatial pattern of TB infections. This targeted approach would save labour and costs to control the spread of the disease.

**Acknowledgements**

I wish to thank all contributing authors for their time and effort in preparing their manuscripts as well as the reviewers whose constructive comments considerably improved all manuscripts. I wish to thank the journal's editor-in-chief George Christakos for his encouragement and support as well as Helen James, from the Journals Editorial Office, for her professionalism in bookkeeping and tracking every manuscript and the associated deadlines with timely reminders.

**References:**

Aanensen, D. M., Huntley, D. M., Feil, E. J., al-Own, F. a., & Spratt, B. G. (2009). EpiCollect: Linking Smartphones to Web Applications for Epidemiology, Ecology and Community Data Collection. *PloS One, 4*, e6968.

Anderson, C. (2008). The end of theory. *Wired magazine, 16*, 16-07.

Andreu-Perez, J., Poon, C. C. Y., Merrifield, R. D., Wong, S. T. C., & Yang, G. Z. (2015). Big Data for Health. *IEEE Journal of Biomedical and Health Informatics, 19*, 1193-1208.

Arah, O. A. (2009). On the relationship between individual and population health. *Medicine, Health Care, and Philosophy, 12*, 235-244.

Artois, J., Lai, S., Feng, L., Jiang, H., Zhou, H., Li, X., Dhingra, M. S., Linard, C., Nicolas, G., Xiao, X., Robinson, T. P., Yu, H., & Gilbert, M. (this volume). H7N9 and H5N1 avian influenza suitability models for China: accounting for new poultry and live-poultry markets distribution data. *Stochastic Environmental Research and Risk Assessment*, 10.1007/s00477-00016-01362-z.

Benson, E. S. (2016). Trackable life: Data, sequence, and organism in movement ecology. *Studies in History and Philosophy of Science Part C: Studies in History and Philosophy of Biological and Biomedical Sciences, 57*, 137-147.

Biggeri, A., Catelan, D., Conesa, D., & Vounatsou, P. (2016). Spatio-temporal statistics: applications in epidemiology, veterinary medicine and ecology. *Geospatial health, 11*.

Biggeri, A., Dreassi, E., Catelan, D., Rinaldi, L., Lagazio, C., & Cringoli, G. (2006). Disease mapping in veterinary epidemiology: a Bayesian geostatistical approach. *Statistical Methods in Medical Research, 15*, 337-352.

Blair, W., & Bar-Shalom, T. (1996). Tracking maneuvering targets with multiple sensors: Does more data always mean better estimates? *IEEE Transactions on Aerospace and Electronic Systems, 32*, 450-456.

Boivin, J., & Ng, S. (2006). Are more data always better for factor analysis? *Journal of Econometrics, 132*, 169-194.

Brunton, L. A., Alexander, N., Wint, W., Ashton, A., & Broughan, J. M. (this volume). Using geographically weighted regression to explore the spatially heterogeneous spread of bovine tuberculosis in England and Wales. *Stochastic Environmental Research and Risk Assessment*, 10.1007/s00477-00016-01320-00479.

Burnham, K. P., & Anderson, D. R. (2002). *Model Selection and Multimodel Inference*. New York: Springer Verlag.

Caggiano, G., Kapetanios, G., & Labhard, V. (2011). Are more data always better for factor analysis? Results for the euro area, the six largest euro area countries and the UK. *Journal of Forecasting, 30*, 736-752.

Chiolero, A. (2013). Big data in epidemiology: too big to fail? *Epidemiology, 24*, 938-939.

Christakos, G. (2010). *Integrative Problem-Solving in a Time of Decadence*: Springer, Netherlands, Dordrecht.

Christakos, G., Zhang, C., & He, J. (this volume). A traveling epidemic model of space–time disease spread. *Stochastic Environmental Research and Risk Assessment*, 10.1007/s00477-00016-01298-00473.

Congdon, P. (this volume). Representing spatial dependence and spatial discontinuity in ecological epidemiology: a scale mixture approach. *Stochastic Environmental Research and Risk Assessment*, 10.1007/s00477-00016-01292-00479.

Demšar, U., Buchin, K., Cagnacci, F., Safi, K., Speckmann, B., Van de Weghe, N., Weiskopf, D., & Weibel, R. (2015). Analysis and visualisation of movement: an interdisciplinary review. *Movement Ecology, 3*, 5.

Dion, E., VanSchalkwyk, L., & Lambin, E. F. (2011). The landscape epidemiology of foot-and-mouth disease in South Africa: A spatially explicit multi-agent simulation. *Ecological Modelling, 222*, 2059-2072.

Donoho, D., & Jin, J. (2015). Higher Criticism for Large-Scale Inference, Especially for Rare and Weak Effects. 1-25.

Enright, J. A., & O'Hare, A. (this volume). Reconstructing disease transmission dynamics from animal movements and test data. *Stochastic Environmental Research and Risk Assessment*, 10.1007/s00477-00016-01354-z.

Evans, M. R., Benton, T. G., Grimm, V., Lessells, C. M., O'Malley, M. A., Moustakas, A., & Weisberg, M. (2014). Data availability and model complexity, generality, and utility: a reply to Lonergan. *Trends In Ecology & Evolution, 29*, 302-303.

Evans, M. R., & Moustakas, A. (2016). A comparison between data requirements and availability for calibrating predictive ecological models for lowland UK woodlands: learning new tricks from old trees. *Ecology and Evolution, 6*, 4812-4822.

Fan, J., Han, F., & Liu, H. (2014). Challenges of Big Data analysis. *National Science Review, 1*, 293-314.

Fei, X., Wu, J., Liu, Q., Ren, Y., & Lou, Z. (2016). Spatiotemporal analysis and risk assessment of thyroid cancer in Hangzhou, China. *Stochastic Environmental Research and Risk Assessment, 30*, 2155-2168.

Ferrè, N., Songyin, Q., Mazzucato, M., Ponzoni, A., Mulatti, P., Morini, M., Fan, J., Xiaofei, L., Shulong, D., Xiangmei, L., & Marangon, S. (2016). GIS Applications to Support Entry-Exit Inspection and Quarantine Activities. In O. Gervasi, B. Murgante, S. Misra, A. M. A. C. Rocha, C. M. Torre, D. Taniar, B. O. Apduhan, E. Stankova & S. Wang (Eds.), *Computational Science and Its Applications -- ICCSA 2016: 16th International Conference, Beijing, China, July 4-7, 2016, Proceedings, Part III* (pp. 85-97). Cham: Springer International Publishing.

Fisher, M. C., Henk, D. A., Briggs, C. J., Brownstein, J. S., Madoff, L. C., McCraw, S. L., & Gurr, S. J. (2012). Emerging fungal threats to animal, plant and ecosystem health. *Nature, 484*, 186-194.

Gange, S. J., & Golub, E. T. (2016). From Smallpox to Big Data: The Next 100 Years of Epidemiologic Methods. *American Journal of Epidemiology, 183*, 423-426.

Gorelick, R. (2011). What is theory? *Ideas in Ecology and Evolution, 4*.

Graham, J. P., Leibler, J. H., Price, L. B., Otte, J. M., Pfeiffer, D. U., Tiensin, T., & Silbergeld, E. K. (2008). The Animal-Human Interface and Infectious Disease in Industrial Food Animal Production: Rethinking Biosecurity and Biocontainment. *Public Health Reports, 123*, 282-299.

Guernier, V., Milinovich, G. J., Santos, M. A. B., Haworth, M., Coleman, G., & Magalhaes, R. J. S. (2016). Use of big data in the surveillance of veterinary diseases: early detection of tick paralysis in companion animals. *Parasites & vectors, 9*, 1.

Heesterbeek, J. (2000). *Mathematical epidemiology of infectious diseases: model building, analysis and interpretation* (Vol. 5): John Wiley & Sons.

Juan, P., Díaz-Avalos, C., Mejía-Domínguez, N. R., & Mateu, J. (in press). Hierarchical spatial modeling of the presence of Chagas disease insect vectors in Argentina. A comparative approach. *Stochastic Environmental Research and Risk Assessment*, 10.1007/s00477-00016-01340-00475.

Kenall, A., Harold, S., & Foote, C. (2014). An open future for ecological and evolutionary data? *BMC Evolutionary Biology, 14*, 66.

Knox, E., & Bartlett, M. (1964). The detection of space-time interactions. *Journal of the Royal Statistical Society. Series C (Applied Statistics), 13*, 25-30.

Krebs, J. R., Anderson, R. M., Clutton-Brock, T., Donnelly, C. A., Frost, S., Morrison, W. I., Woodroffe, R., & Young, D. (1998). Badgers and Bovine TB: Conflicts Between Conservation and Health. *Science, 279*, 817-818.

Lange, M., & Thulke, H.-H. (this volume). Elucidating transmission parameters of African swine fever through wild boar carcasses by combining spatio-temporal notification data and agent-based modelling. *Stochastic Environmental Research and Risk Assessment*, 10.1007/s00477-00016-01358-00478.

Levallois, C., Steinmetz, S., & Wouters, P. (2013). Sloppy data floods or precise social science methodologies? Dilemmas in the transition to data-intensive research in sociology and economics (Chapter 5). In P. Wouters, A. Beaulieu, A. Scharnhorst & S. Wyatt (Eds.), *Virtual Knowledge* MIT Press.

Lonergan, M. (2014). Data availability constrains model complexity, generality, and utility: a response to Evans et al. *Trends In Ecology & Evolution, 29*, 301-302.

Lowe, R., Cazelles, B., Paul, R., & Rodó, X. (in press). Quantifying the added value of climate information in a spatio-temporal dengue model. *Stochastic Environmental Research and Risk Assessment*, 10.1007/s00477-00015-01053-00471.

Lynch, S. M., & Moore, J. H. (2016). A call for biological data mining approaches in epidemiology. *BioData Mining, 9*, 1.

Malesios, C., Kostoulas, P., Dadousis, K., & Demiris, N. (this volume). An early warning indicator for monitoring infectious animal diseases and its application in the case of a sheep pox epidemic. *Stochastic Environmental Research and Risk Assessment*, 10.1007/s00477-00016-01316-00475.

Markatou, M., & Ball, R. (2014). A pattern discovery framework for adverse event evaluation and inference in spontaneous reporting systems. *Statistical Analysis and Data Mining: The ASA Data Science Journal, 7*, 352-367.

Marx, C., Mühlbauer, V., Krebs, P., & Kuehn, V. (2015). Species-related risk assessment of antibiotics using the probability distribution of long-term toxicity data as weighting function: a case study. *Stochastic Environmental Research and Risk Assessment, 29*, 2073-2085.

Maslin, M., & Austin, P. (2012). Uncertainty: Climate models at their limit? *Nature, 486*, 183-184.

Mayer-Schönberger, V., & Cukier, K. (2013). *Big data: A revolution that will transform how we live, work, and think*: Houghton Mifflin Harcourt.

McAllister, J. W. (1996). The evidential significance of thought experiment in science. *Studies In History and Philosophy of Science Part A, 27*, 233-250.

McCormick, T. H., Ferrell, R., Karr, A. F., & Ryan, P. B. (2014). Big data, big results: Knowledge discovery in output from large-scale analytics. *Statistical Analysis and Data Mining: The ASA Data Science Journal, 7*, 404-412.

Michener, W. K. (2015). Ecological data sharing. *Ecological Informatics, 29, Part 1*, 33-44.

Miller, J. A. (2012). Using spatially explicit simulated data to analyze animal interactions: a case study with brown hyenas in northern Botswana. *Transactions in GIS, 16*, 271-291.

Mooney, S. J., Westreich, D. J., & El-Sayed, A. M. (2015). Epidemiology in the Era of Big Data. *Epidemiology (Cambridge, Mass.), 26*, 390-394.

Moustakas, A. (2016). The effects of marine protected areas over time and species' dispersal potential: a quantitative conservation conflict attempt. *Web Ecology, 16*, 113-122.

Moustakas, A., & Evans, M. (2015). Coupling models of cattle and farms with models of badgers for predicting the dynamics of bovine tuberculosis (TB). *Stochastic Environmental Research and Risk Assessment, 29*, 623-635.

Moustakas, A., & Evans, M. R. (2016). Regional and temporal characteristics of bovine tuberculosis of cattle in Great Britain. *Stochastic Environmental Research and Risk Assessment, 30*, 989-1003.

Moustakas, A., & Evans, M. R. (this volume). A big-data spatial, temporal and network analysis of bovine tuberculosis between wildlife (badgers) and cattle. *Stochastic Environmental Research and Risk Assessment*, 10.1007/s00477-00016-01311-x.

Najafabadi, M. M., Villanustre, F., Khoshgoftaar, T. M., Seliya, N., Wald, R., & Muharemagic, E. (2015). Deep learning applications and challenges in big data analytics. *Journal of Big Data, 2*, 1-21.

Nelson, J. C., Shortreed, S. M., Yu, O., Peterson, D., Baxter, R., Fireman, B., Lewis, N., McClure, D., Weintraub, E., Xu, S., Jackson, L. A., & on behalf of the Vaccine Safety Datalink, p. (2014). Integrating database knowledge and epidemiological design to improve the implementation of data mining methods that evaluate vaccine safety in large healthcare databases. *Statistical Analysis and Data Mining: The ASA Data Science Journal, 7*, 337-351.

Nobert, B. R., Merrill, E. H., Pybus, M. J., Bollinger, T. K., & Hwang, Y. T. (2016). Landscape connectivity predicts chronic wasting disease risk in Canada. *Journal Of Applied Ecology, 53*, 1450-1459.

Norman, S. A., Huggins, J., Carpenter, T. E., Case, J. T., Lambourn, D. M., Rice, J., Calambokidis, J., Gaydos, J. K., Hanson, M. B., & Duffield, D. A. (2012). The application of GIS and spatiotemporal analyses to investigations of unusual marine mammal strandings and mortality events. *Marine Mammal Science, 28*, E251-E266.

Oleś, K., Gudowska-Nowak, E., & Kleczkowski, A. (2012). Understanding Disease Control: Influence of Epidemiological and Economic Factors. *PloS One, 7*, e36026.

Ortiz-Pelaez, A., Pfeiffer, D., Soares-Magalhaes, R., & Guitian, F. (2006). Use of social network analysis to characterize the pattern of animal movements in the initial phases of the 2001 foot and mouth disease (FMD) epidemic in the UK. *Preventive Veterinary Medicine, 76*, 40-55.

Pearl, J. (1987). Evidential reasoning using stochastic simulation of causal models. *Artificial Intelligence, 32*, 245-257.

Perretti, C. T., Munch, S. B., & Sugihara, G. (2013). Model-free forecasting outperforms the correct mechanistic model for simulated and experimental data. *Proceedings of the National Academy of Sciences, 110*, 5253-5257.

Pfeiffer, D. U., & Stevens, K. B. (2015). Spatial and temporal epidemiological analysis in the Big Data era. *Preventive Veterinary Medicine, 122*, 213-220.

Picado, A., Guitian, F., & Pfeiffer, D. (2007). Space–time interaction as an indicator of local spread during the 2001 FMD outbreak in the UK. *Preventive Veterinary Medicine, 79*, 3-19.

Piwowar, H. A., & Vision, T. J. (2013). Data reuse and the open data citation advantage. *PeerJ, 1*, e175.

R Development Core Team. (2016). R: A language and environment for statistical computing. R Foundation for Statistical Computing, Vienna, Austria. *ISBN 3-900051-07-0.*

Riad, M. H., Scoglio, C. M., McVey, D. S., & Cohnstaedt, L. W. (this volume). An individual-level network model for a hypothetical outbreak of Japanese encephalitis in the USA. *Stochastic Environmental Research and Risk Assessment*, 10.1007/s00477-00016-01353-00470.

Silver, N. (2012). *The signal and the noise: Why so many predictions fail-but some don't*: Penguin.

Smith?, R., Lee, B. Y., Moustakas, A., Zeigler, A., Prague, l., Santos, R., Chung, M., Gras, R., Forbes, V., Borg, S., Comans, T., Ma, Y., Punt, N., Jusko, W., Brotz, L., & Hyder, A. (2016). Population modelling by examples ii. In *Proceedings of the Summer Computer Simulation Conference* (pp. 1-8). Montreal, Quebec, Canada: Society for Computer Simulation International.

Snow, J. (1855). *On the mode of communication of cholera*: John Churchill.

Toh, S., & Platt, R. (2013). Big data in epidemiology: too big to fail? *Epidemiology, 24*, 939.

Tomley, F. M., & Shirley, M. W. (2009). Livestock infectious diseases and zoonoses. *Philosophical Transactions of the Royal Society B: Biological Sciences, 364*, 2637-2642.

Ward, M. P., & Carpenter, T. E. (2000). Techniques for analysis of disease clustering in space and in time in veterinary epidemiology. *Preventive Veterinary Medicine, 45*, 257-284.

Webb, C. R. (2005). Farm animal networks: unraveling the contact structure of the British sheep population. *Preventive Veterinary Medicine, 68*, 3-17.

Zhang, Z., Chen, D., Liu, W., Racine, J. S., Ong, S., Chen, Y., Zhao, G., & Jiang, Q. (2011). Nonparametric Evaluation of Dynamic Disease Risk: A Spatio-Temporal Kernel Approach. *PloS One, 6*, e17381.